\documentclass[reprint,nofootinbib,showpacs,showkeys,prd]{revtex4-1}
\usepackage{graphicx}
\usepackage{epsfig}
\usepackage{amsmath}
\usepackage{amssymb}
\usepackage{calc}
\begin{document}
\title{The associahedron as a holographic entanglement polytope}
\author{P\'eter L\'evay}
\affiliation{MTA-BME Quantum Dynamics and Correlations Research Group, Department of Theoretical Physics,
Budapest University of  Technology and Economics, 1521 Budapest, Hungary}
\date{\today}
\begin{abstract}
By employing the ${\rm AdS}_3/{\rm CFT}_2$ correspondence in this note we observe an analogy between the structures found in connection with the Arkani-Hamed-Bai-He-Yan (ABHY) associahedron used for understanding scattering amplitudes,
and the one used for understanding space-time emerging from patterns of entanglement. 
The analogy suggests the natural interpretation for the associahedron as a holographic entanglement polytope associated to the ${\rm CFT}_2$ vacuum. 
Our observations hint at the possibility that the factorization properties of scattering amplitudes are connected to the notion of separability of space-time as used in the theory of holographic quantum entanglement. 

\end{abstract}

\maketitle
\section{Introduction}
Using the language of cluster algebras\cite{FZ} in our previous papers\cite{Levay,LB}, in the static setup of ${\rm AdS}_3/{\rm CFT}_2$, we elucidated how patterns of entanglement of the CFT vacuum  manifest themselves in causal patterns of kinematic space. As is well-known\cite{Czech1} after taking a static slice $\mathbb D$ of ${\rm AdS}_3$ kinematic space $\mathbb K$ is the space of geodesics of $\mathbb D$. It turns out that $\mathbb K$ is just $1+1$ dimensional de Sitter space ${\rm dS}_2$ equipped with a two-from. This so called Crofton form is coming from the entanglement entropy calculated for boundary subregions of the CFT vacuum.
One can either regard $\mathbb K$ as an intermediary between bulk and boundary descriptions\cite{Czech1,Czech1b,Czechmera} or as an emergent space-time in its own right\cite{Myers,BoerMyers,Zukowski}.

In \cite{Levay,LB}  
we studied partitions of the boundary $\partial\mathbb D$ into $N$ subregions and the corresponding set of full triangulations of geodesic $N$-gons. This set gives rise to a set of causal patterns of $\mathbb K$ made of causal diamonds. 
O:wqur motivation for these elaborations was the hope to understand the structure of entanglement patterns of the CFT vacuum in terms of causal patterns of $\mathbb K$. It is known\cite{Czech1} that the area labels of the causal diamonds, serving as building blocks for such patterns, encode conditional mutual information associated with the $N$ boundary regions. Moreover, such labels can be related\cite{Levay} to cluster variables of an $A_{N-3}$ cluster algebra.   

It turns out\cite{LB} that for fixed $N$ the structure of the space of causal patterns is related to the structure of the associahedron\cite{Stasheff} ${\mathcal K}^{N-3}$, an $N-3$ dimensional polytope which is well-known to string theorists\cite{Devadoss,Hanson}. 
On this space of causal patterns cluster dynamics acts by a recursion provided by a Zamolodchikov's $Y$-system\cite{Zamo,FrenkelSzenes,Gliozzi} of type $(A_{N-3},A_1)$. We also observed that the space of causal patterns is equipped with a partial order, isomorphic to the Tamari lattice\cite{Tamari},
and the mutation of causal patterns can be encapsulated in a walk of $N-3$ particles interacting in a peculiar manner inside the past light cone of an arbitrarily chosen point on the boundary of ${\rm dS}_2$.

Interestingly precisely the same structures show up\cite{Mizera,AHST} in a well-known but seemingly unrelated research topic concerning the ABHY construction\cite{Nima1} of scattering amplitudes in bi-adjoint $\phi^3$ theory.
The research in this field has begun by formulating scattering amplitudes in planar ${\mathcal N}=4$ SYM in terms of a certain differential form in momentum twistor space\cite{Hodges} culminating in the appearance of the famous notion of the amplituhedron\cite{NimaTrnka} geometrizing their factorization properties.
The ABHY construction can be regarded as the much simpler incarnation of these ideas within the context of the bi-adjoint $\phi^3$ theory\cite{Nima1}. This simpler setting has given rise to another appearance in physics of the associahedron, this time as the amplituhedron for $N$-point tree level amplitudes.
Note that the connection between bi-adjoint scalar amplitudes and associahedra was first given in \cite{Mizera}, where it has also been pointed out that this polytope serves as a twisted cycle. Moreover, the amplitudes then show up as intersection numbers of such cycles.

In this scattering scenario the notion of a kinematic space appeared as well. However, unlike in the entanglement context here this terminology refers to the various ways of labeling the on-shell data of the scattering process, i.e. it is the space of Mandelstam invariants. For tree-level scattering described by $N$ massless momenta this space is an $N(N-3)/2$ dimensional space where the $N-3$ dimensional polytope the kinematic associahedron $\mathcal K^{N-3}$ lives. 

In their recent paper Arkani-Hamed et.al\cite{AHST} posed the question whether there exists a structure in their kinematic space which makes the ABHY associahedron encapsulating the magic of the factorization properties of the scattering amplitudes obvious and inevitable. Their answer is that such a structure exists and it is related to the causal structure of a $1+1$ dimensional space-time
for which they coin the term "kinematic space-time". 
 
After realizing that in our entanglement elaborations we have came accross the very same structures, in a note of our recent paper\cite{LB} we referred to the possibility of relating such a "kinematic space-time" to $\mathbb K$ aka $1+1$ dimensional de Sitter space-time. This note can be considered as a step towards developing the details of an interesting analogy between these two seemingly unrelated strands of research.

The basic message of our note is that 
the associahedron can also be regarded as a holographic entanglement polytope associated to the ${\rm CFT}_2$ vacuum.
Recall that the notion of entanglement polytopes has originally shown up in the literature of quantum information\cite{entpoly} connected to the $N$-representability problem of quantum chemistry\cite{Borland,Klyachko}. Entanglement polytopes are geometric objects\cite{sawicki} capable of witnessing multipartite entanglement via using single particle information contained in the spectra of the one-particle reduced density matrix. In this approach one can associate to an entangled state a convex polytope hence the name, enanglement polytope.

In the holographic context since the influential paper of Ref.\cite{Ooguri} the developement of this idea is also under intense scrutiny\cite{HubenyFort,HubenyArrange,DongCzech,HubenySuperbalance}.
Here the geometric framework of holographic theories allows to analyze the usual entropic inequalities of quantum information in a new context.
The new set of holographic inequalities used here are not universally true for any state, instead they hold for the restricted set of CFT states which have a geometric bulk dual. 
The simplest inequality of that kind is the one\cite{Hayden} which expresses the fact that holographic mutual information is monogamous.

To the best of our knowledge untill now there seem to be no record on the connection between entanglement polytopes and generalized associahedra\cite{FZ,GenerAssoc}. What one naively expects is that to certain CFT states (yet to be identified) such objects are encoding patterns of entanglement of the state in a geometric manner.
In particular the results of this paper show that the associahedron, geometrizing the entanglement patterns of the CFT vacuum of the static ${\rm AdS}_3/{\rm CFT}_2$ setup, can be regarded as an example of such a polytope.

Our elaborations also show that the factorization properties of scattering amplitudes in the ABHY context are analogous to the notion of separability as used in the theory of holographic quantum entanglement. Indeed, factorization in the ABHY setting was shown to be connected to factorization of kinematic space-time in Ref.\cite{AHST}. On the other hand within the framework of the AdS/CFT correspondence it is well-known that space-time factorization can also be regarded as a manifestation for the lack of entanglement\cite{Raam3} i.e. separability.
Combining these two pieces of knowledge and the existing analogy between the findings of Refs.\cite{LB} and \cite{AHST} in this note we are presenting some elementary observations supporting the statement that the analogy is probably deeper.

The organization of this paper is as follows.
In three different subsections of section II. we recapitulate the concept of kinematic space as used in the holographic entanglement context, then the idea of kinematic space time as used in the ABHY scattering one, and then we connect these notions.
In Section III. we observe a formal analogy between entanglement entropies and planar variables which are the propagators showing up in planar graphs\cite{Nima1}. In arriving at this analogy we use the notions of the regularized entropy of Ref.\cite{Casini}, and the gauge degree of freedom associated to the space of horocycles playing the role of the space of regulators\cite{Levay}. 

The subsections of Section IV. contain different elaborations on the idea how an associahedron denoted by $\tilde{\mathcal K}^{N-3}$ is encapsulating entanglement information of the CFT vacuum.
This associahedron is the analogue of the kinematic associahedron, i.e. the ABHY one ${\mathcal K}^{N-3}$ of Ref.\cite{Nima1}.
In the first subsection of IV. we show how a Zamolodchikov $Y$-system of type $(A_{N-3},A_1)$, found in the entanglement context\cite{LB}, can be related to the binary geometries studied in \cite{Nima2} in the scattering setup.
Here we see that the decoupling of incompatible $U$-variables\cite{Nima2} is connected to the vanishing of a certain subset of gauge invariant entanglement measures. This process is displayed as the factorization of a facet of $\tilde{\mathcal K}^{N-3}$.
In the second subsection of IV. we show that the act of gauging away nontrivial entanglement entropies is related to the notion of separability which in turn manifests itself in space-time factorization. Here by space-time we mean a fundamental domain of kinematic space, which is isomorphic to elliptic de Sitter space $dS_2/{\mathbb Z}_2$ as used in Ref.\cite{Verlinde}.
The third subsection of IV. is devoted to a detailed illustration of the result that the associahedron ${\mathcal K}^3$ can be regarded as a holographic entanglement polytope. This result is easy to generalize for arbitrary $N$ yielding the same interpretation for ${\mathcal K}^{N-3}$.
The last subsection of section IV. establishes a connection between our associahedron regarded as an entanglement polytope and the world-sheet associahedron well-known from string theory\cite{Devadoss, Hanson,Cruz}.
 Our conclusions and some comments are left for Section V.

\section{Kinematic space}
\subsection{Kinematic space in ${\rm AdS}_3/{\rm CFT}_2$}

We start by clarifying the meaning of kinematic space in the two different contexts. First in the holographic entanglement, then in the ABHY one.

${\rm AdS}_3$ can be regarded as the hyperboloid in ${\mathbb R}^{2,2}$
characterized by the equation
\begin{equation}
-y_{-1}^2 -y_0^2+y_1^2+y_2^2=-{\ell}^2_{\rm AdS}
\label{ads3}
\end{equation}
where $\ell_{\rm AdS}$ is the AdS radius.  In the following we fix this radius by
$\ell_{\rm AdS}=1$.
Taking the static slice means choosing $y_{-1}=0$.
After stereographic projection from the point $(y_0,y_1,y_2)=(-1,0,0)$ of ${\mathbb H}$ 
(defined by $-y_0^2+y_1^2+y_2^2=-1$ and parametrized by the analogues of polar coordinates $\varphi\in[0,2\pi)$ and $\varrho\in[0,\infty)$)
to the Poincar\'e disk ${\mathbb D}$ lying in the plane $y_0=0$ we obtain the coordinates
\begin{equation}
z=\tanh({\varrho}/2)e^{i{\varphi}}=\frac{y_1+iy_2}{1+y_0}\in{\mathbb D}.
\label{ze}
\end{equation}
An alternative set of coordinates can be obtained by transforming to the upper half plane $\mathbb U$ by a Cayley transformation
\begin{equation}
\tau=i\frac{1+z}{1-z}=\frac{i-y_2}{y_0-y_1}=\xi+i\eta\in{\mathbb U},\qquad \eta>0.
\label{tau}
\end{equation}
In our static considerations we will be referring to the spaces $\mathbb D$ and $\mathbb U$ as the bulk, and the unit circle $S^1\simeq\partial\mathbb D$ and the compactified real line ${\mathbb R}{\mathbb P}^1=\mathbb R\cup\{\infty\}\simeq \partial\mathbb U$ as the boundary.

The geodesics on $\mathbb D$ are given by the formula
\begin{equation}
\tanh\varrho\cos(\varphi-\theta)=\cos\alpha.
\label{geo}
\end{equation}
They are circular arcs in the bulk starting and ending on the boundary. 
Here the extra parameters $\theta\in[0,2\pi]$ and $\alpha\in [0,\pi]$ are labelling the geodesics.
Depicted on the disk $\mathbb D$ the coordinate $\theta$ is the center and $\alpha$ is half the opening angle of the geodesic.
Pairs of geodesics differing in orientation are related by
\begin{equation}
 \theta\leftrightarrow \theta+\pi,\qquad \alpha\leftrightarrow \pi-\alpha.
\label{mobi}
\end{equation}
The space of geodesics labelled by the coordinates $(\alpha, \theta)$ is called the kinematic space\cite{Czech1}.
Topologically the kinematic space $\mathbb K$ is the single sheeted hyperboloid $SO(2,1)/SO(1,1)$ which is the de Sitter space ${\rm dS}_2$.
The induced metric in $\mathbb K$ is of the form
\begin{equation}
ds^2_{\mathbb K}=\frac{d\theta^2-d\alpha^2}{\sin^2\alpha}=\frac{dudv}{\sin^2{\frac{v-u}{2}}}
\label{kinmetric}
\end{equation}
where one can also use the coordinates $(u,v)$ related to the pair $(\alpha,\theta)$
as
\begin{equation}
u=\theta-\alpha,\qquad v=\theta +\alpha.
\label{uv}
\end{equation}
The points $e^{iu},e^{iv}\in\partial{\mathbb D}$ can then be regarded as the starting and the endpoints of a geodesic.
If $\alpha$ is regarded as a time-like coordinate and $\theta$ as a space-like one, then the pair of coordinates $(u,v)$ can be regarded as light cone coordinates.

The space of unoriented geodesics (geodesics up to orientation) is obtained after factoring out with the identification (\ref{mobi}). Since de Sitter space $dS_2$ with ${\ell}_{\rm dS}=1$ can be regarded as the locus of points in ${\bf y}\in{\mathbb R}^{2,1}$ satisfying $y_1^2+y_2^2-y_0^2=1$ 
using the parametrization
\begin{equation*}
y_0=\cot\alpha,\quad y_1=\frac{\cos\theta}{\sin\alpha},\quad y_2=\frac{\sin\theta}{\sin\alpha}
\end{equation*}
one can see that the identification (\ref{mobi}) corresponds to the one of ${\bf y}\sim -{\bf y}$. 
We will refer to this ${\mathbb Z}_2$ identification, or alternatively the one based on (\ref{mobi}), as the M\"obius identification.
Hence the space of unoriented geodesics is ${\rm dS}_2/{\mathbb Z}_2$. Following Ref.\cite{Verlinde} this space will be called elliptic de Sitter space.

 Recall that according to Ref.\cite{Czech1} boundary regions are organized according to a causal structure based on the containment relation.
 This structure gives rise to a natural causal structure for bulk geodesics and their representative points in kinematic space.
Two points in kinematic space are time-like separated when their corresponding geodesics contain each other have no intersection and have the same orientation, i.e. when the corresponding boundary intervals are embedded.
Two points are null separated when their geodesics, hence their corresponding boundary regions, have a common endpoint.
Finally they are spacelike separated when their corresponding geodesics either have intersection or have different orientation without intersection\cite{Czech1,Zhang}.

Finally we recall that the area form associated to the metric of Eq.(\ref{kinmetric}) is related to the Crofton form $\omega$ on kinematic space.
More precisely one has\cite{Czech1}
\begin{equation}
\omega=\frac{\partial^2S(u,v)}{\partial u\partial v}du\wedge dv=\frac{\mathfrak{c}}{12}\frac{du\wedge dv}{\sin^2\left(\frac{v-u}{2}\right)}
\label{Crofton}
\end{equation}
where
\begin{equation}
S(u,v)=\frac{\mathfrak{c}}{3}\log\left(e^{\Lambda}\sin\left(\frac{v-u}{2}\right)\right)
\label{Suv}
\end{equation}
is the entanglement entropy of a boundary region with starting (end) points $e^{iu}$ ($e^{iv}$) on the boundary.
Here $\mathfrak{c}$ is the central charge of the boundary CFT related to the bulk Newton constant $G_{\rm N}$ and the AdS length scale $\ell_{\rm AdS}$ via the Brown-Henneaux\cite{BH} relation: $\mathfrak{c}=\frac{3\ell_{\rm AdS}}{2G_{\rm N}}$.
In the considerations of Section III. an important role will be given to the degree of freedom provided by the choice for the cutoff factor $e^{\Lambda}$.

As shown in\cite{Czech1} this holographic scenario relating the entanglement structure of the vacuum to the geometry of pure $AdS_3$ can substantially be generalized.
There it was argued that for {\it every} static holographic spacetime in the $AdS_3/CFT_2$ correspondence it is possible to choose a Crofton form $\omega$ playing the role as the measure
on the space of geodesics. This measure makes it possible to calculate the length of a curve in the bulk via integrating for the geodesics intersecting the curve\cite{Czech1}.
The correct choice for the measure $\omega$ in the space of geodesics is
of the same form as the one that can be found on the left hand side of Eq.(\ref{Crofton}) with $S(u,v)$ replaced by the entanglement entropy of the interval $(u,v)$ as calculated by the Ryu-Takayanagi proposal\cite{RT,RT2}.
This Crofton form can alternatively be used as an area form on the corresponding $\mathbb K$ kinematic space.

Having discusse the notion of kinematic space in the entanglement context, let us now turn to kinematic space as defined by ABHY\cite{Nima1}.

\subsection{Kinematic space in the ABHY setting}

In the ABHY context kinematic space for $N$ massless momenta $p_a$, $a=0,1,\dots,N-1$  is the space spanned by the linearly independent Mandelstam invariants (see Refs. \cite{AHST} and \cite{Nima1} for more details)
\begin{equation}
s_{ab}=(p_a+p_b)^2=2p_a\cdot p_b.
\label{Mandelstam}
\end{equation}
Notice that due to the massless on shell condition and momentum conservation 
$\sum_{a=0}^{N-1}p_a=0$,  we have a polygon in momentum space with vertices at $x_a$ where $p_a=x_{a+1}-x_{a}$ and $x_N=x_0$.
Due to momentum conservation 
we have $\sum_as_{ab}=0$ for all $b=0,1,\dots N-1$ hence
the dimension of kinematic space is $N(N-3)/2$. 
We will refer to the ABHY kinematic space as $K^{N(N-3)/2}$.

We visualize the particle labels $a=0,1,\dots,N-1$ as the ones oredered cyclically in a counter clock-wise manner on the unit circle.
Then one can define planar variables 
\begin{equation}
X_{a,b}=(p_a+\dots +p_{b-1})^2=(x_b-x_a)^2
\label{Nimaentropy}
\end{equation}
which are the propagators appearing  in planar graphs. Notice also that 
\begin{equation}
X_{a,b}=X_{b,a}.
\label{komplementer}
\end{equation}
A further important point to be recorded is that
\begin{equation}
X_{a,a+1}=0.
\label{vanishing}
\end{equation}
So the not necessarily zero $X_{a,b}$s correspond to non adjacent pairs $(a,b)$ of the cyclically ordered particles, and their number coincides with the dimension of kinematic space.
As a result of this the Mandelstam variables can be expanded in terms of the planar variables 
\begin{equation}
s_{ab}=X_{a,b+1}+X_{a+1,b}-X_{a,b}-X_{a+1,b+1}.
\label{presubadditiv}
\end{equation}

The ABHY associahedron is a polytope that lives in $K^{N(N-3)/2}$. For its construction the authors of \cite{Nima1} first define a simplex $\Delta^{N(N-3)/2}$ in kinematic space by imposing the constraints
\begin{equation}
X_{a,b}\geq 0
\label{entropypositive}
\end{equation}
with $0\leq a<b\leq N-1$.
 
Then for a set of constants $c_{a,b}\geq 0$, 
for every pair of {\it non-adjacent} indices $0\leq a<b\leq N-2$,
they consider the set of additional constraints
$c_{a,b}=-s_{a,b}$. This means that we also have the constraint 
\begin{equation}
c_{a,b}=X_{a,b}+X_{a+1,b+1}-X_{a,b+1}-X_{a+1,b}\geq 0.
\label{subadd}
\end{equation}
Eq.(\ref{subadd}) results in a restriction to an $N-3$ dimensional subspace $H^{N-3}$ of kinematic space.
Using then both of our (\ref{entropypositive}), (\ref{subadd}) constraints , the associahedron ${\mathcal K}^{N-3}$ shows up as $\Delta^{N(N-3)/2}\cap H^{N-3}={\mathcal K}^{N-3}\subset K^{N(N-3)/2}$.

The aim of this formalism as developed by ABHY was to make structures of basic importance of scattering amplitudes in bi-adjoint $\phi^3$ theory explicit, ones that have been hidden in the usual Feynmann diagram treatment building up on the principles of locality and unitarity. Tree level amplitudes have poles when a subset of momenta goes on-shell. In this case the corresponding residues on the pole factorize. However, apart from this well-known phenomenon, there is also the one of certain patterns of poles appearing together. It has turned out that for a fixed $N$ such compatible poles showing up together correspond to the vertices of ${\mathcal K}^{N-3}$. Moreover, the factorization properties of ${\mathcal K}^{N-3}$ to lower dimensional associahedra then automatically take care of issues of factorizability of the amplitudes.

The main motivation of Ref.\cite{AHST} was to find an extra structure which makes both factorization and the compatibility structures of poles self evident. Then the authors show that the  answer to this question lies in the causal structure of a $1+1$ dimensional space-time equipped with a notion of "positivity".
This space-time is defined as a continuum limit of a discretized space showing up in the labels of the planar variables $X_{a,b}$. Here the two labels for time and space are related to the indices $a,b$. They emphasize that the index $a$ is also enjoying a cyclic symmetry $a\mapsto a+N$ which dates back to the cyclic arrangement of our momenta $p_a$.
They call this structure {\it kinematic space-time}.

\subsection{Connecting kinematic spacetime to $\mathbb K$}

In this subsection we observe that the kinematic space-time of Ref.\cite{AHST}, suitably extended, corresponds to $\mathbb K$ with the roles of space and time exchanged.

In order to see this just have a look at Eq.(\ref{geo}) and notice that  points lying on the boundary ($\varrho\to\infty$) of $\mathbb D$ define point curves\cite{Czech1} in $\mathbb K$. These point curves give rise to a rectangular grid 
of light cone coordinates
precisely of the form seen in  Figure 2. of Ref.\cite{AHST}.  
Indeed, the analogues of the labels $(a,b)$ used for the planar variables $X_{a,b}$ are just discretized ones of the light cone coordinates $(u,v)$ of Eq.(\ref{uv}). Moreover, the periodicity under $a\to a+N$ for one of the labels is just the discretized version of the periodicity of $\theta$.

Now a pair $(a,b)$ from our cyclically arranged particle labels is associated with a pair of boundary points, which in turn determines a geodesic of the bulk up to orientation. Indeed, to a pair of boundary points one can associate two geodesics with opposite orientation.  Hence if we consider merely the space of geodesics up to orientation one should also take into account the M\"obius identification of points as described in \cite{AHST}. 
Comparing Figure 2. of Ref.\cite{AHST} labelled by the coordinates $(t,x)$ and the right hand side of Figure 4. of Ref.\cite{LB} labelled by the ones of $(\alpha,\theta)$ the analogy between kinematic space-time and $\mathbb K$ is obvious, with two important caveats.

The first is that the roles of space and time like coordinates in the two different settings are exchanged. The second is that Figure 2. of Ref.\cite{AHST} merely reproduces a part of Figure 9. of Ref.\cite{LB}, that part which we called the central belt of kinematic space. In order to obtain a complete correspondence, kinematical spacetime has to be extended. This amounts to a formal inclusion of extra grid points to the left and right hand sides of Figure 2. of Ref.\cite{AHST} resulting in the appearance of extra causal diamonds. As we see in Section IV.A. their trivially vanishing cluster variables provide the correct boundary conditions for a Zamoldochikov Y-system.

Now the scattering picture (see Figure 5. of \cite{Nima1}) is simply related to the geodesic picture with the geodesics playing the roles of diagonals of some (partial) triangulation of $\mathbb D$.
Recall also that boundary regions enjoy the causal ordering mentioned in subsection A. In particular two boundary regions are space-like separated if the corresponding geodesics cross. In $\mathbb K$ this corresponds to space-like separation (time-like separation in \cite{AHST}) between the corresponding points representing these geodesics, with respect to the metric of Eq.(\ref{kinmetric}) conformally equivalent to the flat metric.
This situation corresponds to crossing diagonals in \cite{Nima1} an important case needed for establishing the associahedron structure (see later).

The upshot of these simple considerations is that kinematic space-time, suitably extended, with the roles of space and time exchanged corresponds to the space of geodesics in $\mathbb D$ i.e. $\mathbb K$ which is ${\rm dS}_2$. With the M\"obius identification of Eq. (\ref{mobi}) it corresponds to elliptic de Sitter space ${\rm dS}_2/{\mathbb Z}_2$.   
One can define  ${\rm dS}_2/{\mathbb Z}_2$ as a {\it fundamental domain} in  ${\rm dS}_2$. 
In the following we will need a description of the discretized version of this domain.
In the discretized version instead of the continuously changing pair $(u,v)$ subject to the  $(u,v)\sim (v,u)$ M\"obius identification, one can choose the light-like coordinate grid characterized by the pair $(a,b)$ with $a,b=0,1,\dots N-1$ provided by point curves. Now fixing a fundamental domain simply means that from the space of unconstrained $(a,b)$ pairs one can restrict attention to the domain of pairs that does not redundantly include both $(a,b)$ and $(b,a)$. 
In this case for a fixed $N$ the fundamental region represents the discretized version of ${\rm dS}_2/{\mathbb Z}_2\simeq {\mathbb K}/{\mathbb Z}_2$ properly.
In the scattering context such domains are similar to the ones depicted in Figure 4. of Ref.\cite{AHST}). Alternatively the red regions that one can see in our Figure 2. can be considered as such domains. They are called {\it causal patterns} in \cite{LB}. In Figure 2. where these patterns are represented in $dS_2$, one can observe that there are always two such patterns which should be M\"obius identified when regarding them as causal patterns of ${\rm dS}_2/{\mathbb Z}_2$.

In \cite{Verlinde} it has been argued that the space  ${\rm dS}_2/{\mathbb Z}_2$ has remarkable properties. Unlike ${\rm dS}_2$ this space has a single spacelike boundary, hence providing better perspectives for a holographic description of its quantum gravity. 
Moreover, this space is capable of implementing observable complementarity.
In the following we are adopting the philosophy to investigate the structure of this space-time as represented  by the entanglement structure of a CFT vacuum on this boundary.
As a first step,
having related kinematic space-time and ${\rm dS}_2/{\mathbb Z}_2$, let us see how to also relate the extra structures these spaces exhibit.

\bigskip

\section{Entanglement entropy and planar variables}

Let us now use the upper half plane model of our bulk
and choose a boundary region $A$ with end points labelled by the coordinates $\xi_b,\xi_c\in {\mathbb R}$ (see Eq.(\ref{tau}) and Figure 1.). 
We denote the bulk geodesic anchored to $A$ by ${\bf A}$.
Since the length of this geodesic diverges we have to regularize it.
For regularization we choose a pair of horocycles $h_b$ and $h_c$ which are circles tangent to the boundary at $\xi_b$ and $\xi_c$.
Then by virtue of the Ryu-Takayanagi formula\cite{RT,RT2,HRT} for the CFT vacuum the {\it regularized}\cite{Casini} entanglement entropy of $A$ can be written as\cite{Levay} 
\begin{equation}
S_A=\frac{\mathfrak{c}}{3}\log \lambda(\bf A).
\label{entropy}
\end{equation}
For regularization we used the lambda length $\lambda(\bf A)$ introduced by Penner\cite{Penner,Pennerbook}. This quantity is related to the regularized geodesic length $d(\tau_b,\tau_c)$ of the geodesic $\bf A$ by the formula
\begin{equation}
\lambda({\bf A})=\frac{L(A)}{\sqrt{\Delta_b\Delta_c}}=e^{d(\tau_b,\tau_c)/2},\qquad L(A)=\vert\xi_c-\xi_b\vert.
\label{simplelambda}
\end{equation}
Here $\Delta_b$ and $\Delta_c$ are the Euclidean lengths of the diameters of the corresponding horocycles, and $\tau_b,\tau_c\in {\mathbb U}$ are the endpoint coordinates of the geodesic segment lying in between the corresponding horocycles.
For more on horocycles, lambda lengths and how they are related to a regularization more familiar from the literature\cite{Maxfield} see Appendix A. of Ref.\cite{Levay}.

\begin{figure}
\centerline{\includegraphics{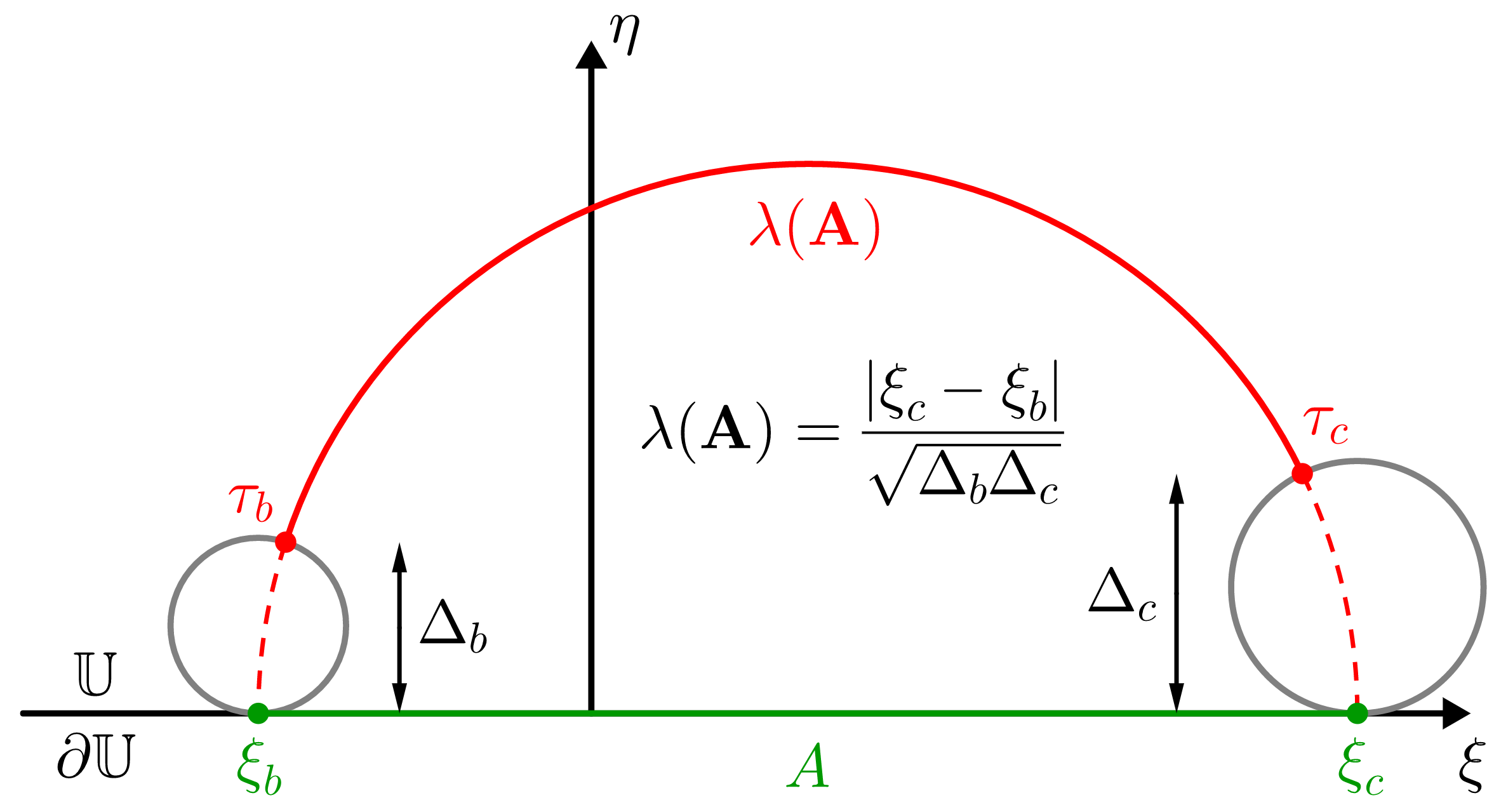}}
\caption{Illustration of the meaning of the lambda length for a circular arc centered on ${\mathbb R}$ which is a part of the boundary $\partial{\mathbb U}$. According to Penner\cite{Penner} the lambda length is the regularized length of the geodesic, where the regulators are horocycles. On $\mathbb U$ these are circles with Euclidean diameters $\Delta$ providing a natural cutoff.}
\end{figure}

There is an important caveat to applying formula (\ref{entropy}).  
Depending on how we choose our horocycles the lambda length can also be unity or less than unity, rendering $S(\bf A)$ zero or negative.
Note that in Ref.\cite{Casini} our $S$, denoted there by $G$, was called the {\it regularized entropy} which is finite and is to be be contrasted with the entanglement entropy which is ultraviolet divergent indicating that it has no covariant meaning. Now, just like our $S$, this regularized entropy of \cite{Casini} can be zero or even negative, but this possibility causes no problem since its physical meaning is only up to a constant term. 
Usually in the literature  horocycles are having infinitesimal euclidean diameters, hence the situation of $S\leq 0$ is avoided by simply refraining from making use of the full set of regulators.
However in Ref.\cite{Levay} we have argued that like $\mathbb K$ the space of geodesics, the space of horocycles is also an important object. Its physical meaning is associated with the gauge degrees of freedom elaborated in \cite{Czech2,Levay}. 
As nicely summarized in \cite{Czech2} in a CFT without a scale setting the cutoff is a gauge choice.
According to this philosophy as the "space of cutoffs" the space of horocycles is to be used, which is a homogeneous space\cite{Levay} like $\mathbb K$.
In the following we argue that the use of this gauge degree of freedom provides the basis for obtaining additional insight on issues of holography, entanglement and the ABHY associahedron.

The most important application of the gauge degree of freedom rests in the possibility of gauging away a certain subset of the entanglement entropies. This possibility is arising when we are interested in the entanglement properties of a large set of boundary regions. This is the case when we have an $N$-fold partition of the boundary with $N$ large, and we are having in mind the possibility of taking the continuum limit later on. 

Let us then consider a sequence of cyclically ordered set of boundary regions $A_0,A_1,\dots ,A_{N-1}$ giving a partition of the boundary. The end points of these regions are cyclically labelled in a counter clock-wise manner by $j=0,1,\dots N-1$. These points $\xi_0,\xi_1,\dots \xi_{N-1}\in \partial{\mathbb U}$, with $A_j\equiv [\xi_{j},\xi_{j+1}]$, are lying on the boundary and they are giving rise to a geodesic $N$-gon with oriented {\it geodesics} ${\bf A}_0,{\bf A}_1,\dots ,{\bf A}_{N-1}$ 
in the bulk ${\mathbb U}$. 
In $\mathbb K$ these geodesics give rise to {\it pairs of points} ${\mathcal A}_0,{\mathcal A}_1,\dots,{\mathcal A}_{N-1}$, and $\overline{\mathcal A}_0,\overline{\mathcal A}_1,\dots,\overline{\mathcal A}_{N-1}$ corresponding to oppositely oriented geodesics.
Using the antipodal map in $\mathbb K/{\mathbb Z}_2$ a point ${\mathcal A}$ and its antipodal version $\overline{\mathcal A}$ is identified. 
In the following we use the notation $A_{a,b}$ and ${\bf A}_{a,b}$ referring to a boundary region and its corresponding bulk geodesic anchored to the points $a$ and $b$. Likewise ${\mathcal A}_{a,b}$ refers to the point of ${\mathbb K}$ lying at the intersection of the point curves labeled by $a$ and $b$. Note that in this notation we have $A_a\equiv A_{a,a+1}$. 

Choosing horocycles conveniently one can achieve\cite{Levay} that  all of the {\it regularized entropies} of such boundary regions are gauged away i.e. we have
\begin{equation}
S({\bf A}_0)=S({\bf A}_1)=\cdots=S({\bf A}_{N-1})=0.
\label{gaugeaway}
\end{equation}
Since the boundary regions are cyclically oriented, due to counter clock-wise orientation of the points, then one can define $S_{a,a+1}\equiv S({\bf A}_a)$. Hence for $a=0,1,\dots,N-1$ Eq.(\ref{gaugeaway}) takes the form
\begin{equation}
S_{a,a+1}=0.
\label{kezdet}
\end{equation}
Now changing the orientation of a geodesic ${\bf A}_a$ results in the new geodesic $\overline{\bf A}_a$  anchored to the boundary region $\overline{A}_a$ complementary to $A_a$. Since the CFT vacuum is a pure state then we have $S(\overline{\bf A}_a)=S({\bf A}_a)$ hence $S_{a,a+1}=S_{a+1,a}=0$. Of course on can also apply this rule for the more general setting of nonvanishing entropies yielding
\begin{equation}
S_{a,b}=S_{b,a}.
\label{folyt1}
\end{equation}

Obviously after choosing in (\ref{simplelambda}) the gauge $\lambda_{a,b}\geq 1$ the entropies are also satisfying
\begin{equation}
S_{a,b}\geq 0.  
\label{posent}
\end{equation}
Since in this case the entropies are nonnegative, using horocycles in this manner can be interpreted as using the {\it physical branch} of the regularized entropies.

After comparing Eqs.(\ref{kezdet})-(\ref{folyt1}) with Eqs.(\ref{komplementer})-(\ref{vanishing}) we see that
there is a formal similarity between the behavior of the planar variables in the ABHY setting and the von Neumann entropies in the holographic entanglement setting provided, we restrict attention to an  "entropy simplex" similar to the simplex $\Delta^{N(N-3)/2}$.

Hence we are led to look at the (formal) analogy
\begin{equation}
{\alpha}^{\prime}X_{a,b}\leftrightarrow \frac{3}{\mathfrak{c}}S_{a,b}
\label{alapmegfeleles}
\end{equation}
where the constant $\alpha^{\prime}$ is of dimension length squared. Then the combination on the left hand side is reminsicent of the exponent used in the Koba-Nielsen formula\cite{KN} as rewritten in Eq.(8) of Ref.\cite{Nima2} (we get back to this point later).

If we are willing to accept the (\ref{alapmegfeleles}) analogy even much more can be said. The von Neumann entropies are satisfying the strong subadditivity constraint\cite{LiebRusskai,NC} 
which for a geodesic quadrangle with $a<b<c<d$ can be rephrased\cite{Levay,LB} as
	\begin{equation}
I(ab,cd\vert bc)\equiv S_{a,c}+S_{b,d}-S_{b,c}-S_{a,d}\geq 0.
\label{strongsubadditivity}
\end{equation}
Here on the left hand side the quantity $I(ab,cd\vert bc)$ is the conditional mutual information\cite{Czech1,NC}.
Unlike the entanglement entropy which is ultraviolet divergent $I(ab,cd\vert bc)$ is finite (unless $b=c$) since the divergences
cancel. Since this quantity is well-defined in \cite{Casini} it has been proposed as the meaningful entanglement entropy.
Strong subadditivity is the statement that the conditional mutual informations are nonnegative.
A reformulation of this statement is connected to the notion of entanglement monogamy\cite{NC}.

Clearly when $a$ and $b$  are non adjacent one obtains
\begin{equation*}
I(aa+1,bb+1\vert a+1b)\equiv S_{a,b}+S_{a+1,b+1}-S_{a+1,b}-S_{a,b+1}\geq 0.
\end{equation*}
Comparing this equation with Eq.(\ref{subadd}) shows that relating the planar variables to entanglement entropies {\it automatically} ensures the validity of Eq.(\ref{subadd}) with the correspondence
\begin{equation}
{\alpha}^{\prime}c_{a,b}\leftrightarrow \frac{3}{\mathfrak{c}}I_{a,b}
\label{fontos1}
\end{equation}
where
\begin{equation}
I_{a,b}\equiv I(aa+1,bb+1\vert a+1b).
\label{atcimkez}
\end{equation}
An alternative way for writing this quantity is
\begin{equation}
\frac{3}{\mathfrak{c}}I_{a,b}=\log(1+t_{a,b})=-\log U_{a,b}
\label{osszefuggesek}
\end{equation}
where
\begin{equation}
U_{a,b}=\frac{(a+1b)(ab+1)}{(a+1b+1)(ab)}\equiv\frac{(\xi_{a+1}-\xi_b)(\xi_a-\xi_{b+1})}{(\xi_{a+1}-\xi_{b+1})
(\xi_a-\xi_b)}.
\label{uvar}
\end{equation}
Here 
$\log t_{a,b}$ is the shear coordinate for the geodesic quadrangle\cite{Pennerbook,Levay} in $\mathbb U$ defined by the four points $\xi_a<\xi_{a+1}<\xi_b<\xi_{b+1}\in \partial{\mathbb U}$.
One can show\cite{Czech1,LB} that Eq.(\ref{osszefuggesek}) encapsulates the fact that $I_{a,b}$ is the area in ${\mathbb K}$, with respect to the (\ref{Crofton}) Crofton form, of the causal diamond defined by the point curves 
of the quadruplet of boundary points  $a,a+1,b,b+1$.
One also has\cite{Zhang,LB}
\begin{equation*}
t_{a,b}=\sinh^2\frac{\Delta\tau_{a,b}}{2}
\end{equation*}
where $\Delta\tau_{a,b}$ is the proper time, calculated with respect to the (\ref{kinmetric}) metric, elapsed between the space-time events associated with the points $\mathcal{A}_{a+1,b}$ (past) and $\mathcal{A}_{a,b+1}$ (future) of the causal diamond.

Notice that after regarding the $I_{u,v}$  for every pair of non-adjacent indices
$0\leq u<v\leq N-2$ as basic variables attached to elementary causal diamonds the (\ref{strongsubadditivity}) general form of the strong subadditivity relation can be written as
\begin{equation*}
I(ab,cd\vert bc)=\sum_{u=a}^{b-1}\sum_{v=c}^{d-1} I_{u,v}=S_{a,c}+S_{b,d}-S_{b,c}-S_{a,d}\geq 0
\end{equation*}
where the first equality expresses the additivity of conditional mutual information\cite{Czech1}. The reader should compare this formula with Eq. (3.8) of \cite{Nima1}.

Recall now from Section II.B that imposing the constraint of Eq.(\ref{subadd}) in the ABHY setting is equivalent to restricting our attention to the subspace $H^{N-3}$. Moreover, the simultaneous implementation of {\it both} of the constraints of Eqs.(\ref{entropypositive}) and (\ref{subadd}) yields\cite{Nima1} the associahedron ${\mathcal K}^{N-3}$.
In the entanglement context the constraint (\ref{posent}) expressing the nonnegativity of the entropies and the (\ref{strongsubadditivity}) one of strong subadditivity are precisely of the same form. Hence by analogy, a similar construction yields another copy of the associahedron $\tilde{\mathcal K}^{N-3}$. A picture
of $\tilde{\mathcal K}^{3}$
 featuring the quantities $S_{a,b}$ and $I_{a,b}$ can be seen in Figure 5.
Using our analogy and the ideas of Ref.(\cite{AHST}) its detailed entanglement based construction will be presented in Section IV.C.
What is the connection between $\tilde{\mathcal K}^{N-3}$ and ${\mathcal K}^{N-3}$?
An elaboration on this important question is postponed to Section IV.D.

For now we stress that strong subadditivity is a highly nontrivial constraint lying at the core of many important physical applications in quantum information.
Here we see that by accepting our analogy it also implements naturally the positivity requirements of the ABHY construction. Notice that in \cite{AHST} the positivity requirements were injected into the kinematical space-time approach by hand. Here the physical properties of entanglement automatically imply such constraints of positivity.

\section{Patterns of entanglement of the CFT vacuum and the associahedron}

\subsection{Zamolodchikov Y-system}

As a first step of elaborating on our analogy using (\ref{atcimkez}) we relabel the data associated with the 
$N(N-3)/2$ causal diamonds in the notation of \cite{LB}. Let
\begin{equation}
I_{j,k}\equiv I_{a-1,b}
\label{areas}
\end{equation}
with $j=1,2,\dots N,\quad k=1,\dots,N-3$
and
\begin{equation}
a\equiv\frac{j-k}{2},\quad b\equiv\frac{j+k}{2}\quad{\rm mod}N.
\label{ujra}
\end{equation}
By virtue of (\ref{osszefuggesek}) let us then define
\begin{equation} U_{j,k}=e^{-\frac{3}{c}
I_{j,k}},\qquad j+k\equiv 0\quad{\rm mod}2.
\label{lenyeg}
\end{equation}

With the change of variables
\begin{equation}
Y_{j,k}=\frac{U_{j,k}}{1-U_{j,k}}
\label{ipszilon}
\end{equation}
the recursion relation\cite{LB}
\begin{equation}
Y_{j-1,k}Y_{j+1,k}=(1+Y_{j,k-1})(1+Y_{j,k+1})
\label{zamoan}
\end{equation}
with the boundary conditions
\begin{equation}
Y_{j,0}=
Y_{j,N-2}=0
\label{bcond}
\end{equation}
holds.
This system is
a Zamolodchikov $Y$ system of type $(A_{N-3},A_1)$.
As is well-known such systems satisfy Zamolodchikov periodicity\cite{Zamo,FrenkelSzenes,Gliozzi} in the form
\begin{equation}
Y_{j+N,N-2-k}=Y_{j,k}.
\label{mobius}
\end{equation}
Notice also that in an obvious notation 
\begin{equation}
Y_{j,k}=\frac{1}{t_{j,k}}.
\label{ipszishear}
\end{equation}

The solution\cite{FrenkelSzenes} of the  $Y$ system is given
in terms of a special set of $U_{j,k}$s which we call a {\it seed set}.
In the notation of Ref.\cite{FrenkelSzenes} let us define
\begin{equation}
U_k\equiv U_{k,k}\qquad k=1,\dots,N-3 
\label{seedset}
\end{equation}
with the boundary condition of Eq.(\ref{bcond}) for $j=0$ and $j=N-2$ translated into $U_0=U_{N-2}=0$. 
Then
\begin{equation}
U_{j+2,j}=\frac{1-U_1U_2\cdots U_j}{1-U_1U_2\cdots U_{j+1}}
\label{fs1}
\end{equation}

\begin{equation}
U_{N-4-j,N-2-j}=\frac{1-U_{N-3}U_{N-4}\cdots U_{N-2-j}}{1-U_{N-3}U_{N-4
}\cdots U_{N-3-j}}.
\label{fs2}
\end{equation}
Note that for the special case of a regular geodesic $N$-gon we have no $j$ dependence and we have the explicit form\cite{LB,FrenkelSzenes}
\begin{equation}
U_{j,k}\equiv 1-\frac{\sin^2\kappa}{\sin^2((k+1)\kappa)},\quad \kappa=\frac{\pi}{N}
\label{jeindep}
\end{equation}
resulting in
\begin{equation}
I_{j,k}=\frac{\mathfrak{c}}{3}\log\frac{\sin^2((k+1)\kappa)}{\sin (k\kappa)\sin((k+2)\kappa)}.
\label{zsolt}
\end{equation}

Now according to Proposition 4. of Ref.\cite{FrenkelSzenes}
\begin{equation}
1-U_{j,k}=\prod_{(j^{\prime},k^{\prime})\in \overline{\mathcal C}_{j,k}}U_{j^{\prime},k^{\prime}}
\label{kupos}
\end{equation}
where $1\leq j\leq N$ and $1\leq k\leq N-3$ , $j+k\equiv 0$ mod$2$ and
\begin{equation*}
\overline{\mathcal C}_{j,k}=\{(j^{\prime},k^{\prime})\vert \vert k^{\prime}-k\vert <\vert j^{\prime}-j\vert\}.
\end{equation*}
The $\overline{\mathcal C}_{j,k}$ symbol defines the complement of the light cone of the point labelled by $(j,k)\in\mathbb K$.

Reverting to the usual labelling by $U_{a,b}$ one can choose the seeds as $U_{N-1,k}$ with $k=1,2,\dots N-3$. Moreover, we realize that these equations give the basic example of {\it binary geometries}  of Ref.\cite{Nima2}, namely one has
\begin{equation}
U_{a,b}+\prod_{(c,d)\in(a,b)^{\times}}U_{c,d}=1
\label{binary}
\end{equation}
where $(c,d)\in(a,b)^{\times}$ refers to the set of all diagonals in $\mathbb D$ that cross the diagonal $(a,b)$. It means that the set of boundary regions $[c,d]$ are space-like separated from the fixed one $[a,b]$. Pairs of points in $\mathbb K$ like $(a,b)$ and $(c,d)$ are called {\it incompatible} in Ref.\cite{Nima2}.

Notice that by virtue of $U_{a,b}\geq 0$ the solution space $\mathcal{U}^+$ of (\ref{binary}) implements positivity. Moreover, the (\ref{binary}) binary relations are also compatible with the condition $0\leq U_{a,b}\leq 1$. We also know that the solution space $\mathcal{U}^+$ is $N-3$ dimensional and spanned by the seed solutions. The (\ref{binary}) relations are called binary relations, since sending $U_{a,b}\to 0$ is rendering the incompatible $U_{c,d}\to 1$. Alternatively, if $U_{a,b}=1$ then we must have $U_{c,d}=0$ for at least one $(c,d)\in (a,b)^{\times}$.

 Using (\ref{osszefuggesek}) $U_{c,d}\to 1$ in  $\mathbb K$ means that for the conditional mutual informations we have $I_{c,d}\to 0$, i.e. the areas of the corresponding causal diamonds in $\mathbb K$ are shrinking to zero. Alternatively one can say that\cite{Casini} in $\partial\mathbb D$ the entanglement between the overlapping regions $[c,d]$ and $[c+1,d+1]$ goes to zero.
On the other hand $U_{a,b}\to 0$ means that the corresponding causal diamond stretches out to the boundary of ${\mathbb K}$ hence its area label ($I_{a,b}$) with respect to the Crofton form goes to infinity. 
Reaching out to the boundary in $\mathbb K$ means that one of the boundary regions in $\partial\mathbb D$ is pinching to a point. 
This is the case for example when $\xi_b\to \xi_{a+1}$ in the (\ref{uvar}) expression of the cross-ratio.
In this case the finite entanglement measure between the overlapping regions $[a,b]$ and $[a+1,b+1]$ will be divergent due to the lack of overlap\cite{Casini}. Since by virtue of Eq.(\ref{binary}) they can be converted to each other
the complementary cases of $I_{a,b}\to 0$ or $I_{a,b}\to\infty$ describe the same important boundary situations.

Let us consider the $U_{ab}\to 0$ case for certainty. From the boundary point of view this case corresponds to the one when $\xi_{a+1}=\dots =\xi_b$ or $\xi_{b+1}=\dots =\xi_0=\xi_1=\dots \xi_a$. From the bulk point of view this means that we have a diagonal $(ab)$ and the points forming a subpolygon on the two sides of this diagonal are pinching together.
For example for $N=6$  one can choose the $N(N-3)/2=9$ plaquettes labelled by the pairs: $02,03,04,13,14,15,24,25,35$.
This choice corresponds to choosing the fundamental domain the past light cone of the point $0$.
Let us now choose the diagonal $14$.
Then under $\xi_2=\xi_3=\xi_4$ our hexagon degenerates to the quadrangle $1450$. Under this we have $U_{1,4}\to 0$ hence $U_{0,2}=U_{0,3}=U_{2,5}=U_{3,5}\to 1$. Then for the entanglement measures we have $I_{0,2}=I_{0,3}=I_{2,5}=I_{3,5}\to 0$ on the other hand $I_{1,3}=I_{2,4}=I_{1,4}$ diverge.
However, $I_{0,4}$ and $I_{1,5}$ stays nonzero.
Similarly under $\xi_5=\xi_0=\xi_1$  the same set stays zero but now $I_{0,4}=I_{1,5}=I_{1,4}$ diverge and $I_{1,3}$ and $I_{2,4}$ stays nonzero. This corresponds to the situation when the hexagon degenerates to the other quadrangle $1234$.
This situation corresponds to the one when the facet $F_{14}$ as a codimension one boundary of $\tilde{\mathcal K}^3$ factorizes 
as: $F_{14}=\partial_{14}\tilde{\mathcal K}^3=\tilde{\mathcal K}^1\times\tilde{\mathcal K}^1$.
Clearly in this picture this factorization is governed by the vanishing of the entanglement measures $I_{0,2}=I_{0,3}=I_{2,5}=I_{3,5}\to 0$.

Note that the usual term for this process is called in \cite{Nima2} as the decoupling of the incompatible $U$-variables.
Here we see that the $U$ variables are related to conditional mutual informations via Eq.(\ref{lenyeg})
and their decoupling is connected to the lack of entanglement i.e. separability.
Moreover for $N$ arbitrary, choosing a geodesic labelled by the pair $(a,b)$ which subdivides the bulk into a geodesic $M$-gon and an $N-M+2$-gon one can write for the corresponding facet of the associahedron\cite{Nima1}
\begin{equation}
F_{ab}=\partial_{ab}\tilde{\mathcal K}^{N-3}=\tilde{\mathcal K}^{M-3}\times\tilde{\mathcal K}^{N-M-1}
\label{ufactor}
\end{equation}
hence this decoupling process via a lack of entanglement is encapsulated in the combinatorial factorization of the associahedron to lower dimensional associahedra.

Finally we remark that in
the entanglement picture the (\ref{mobius}) Zamolodchikov periodicity is the discretized version of the antipodal map on ${\mathbb K}$ which is a space-time equipped with a metric related to the Crofton form. As we noted this antipodal map has already been used for defining the notion of elliptic de Sitter space\cite{Verlinde}.
One can regard this space as a spacetime in its own right or as the space of unoriented geodesics in $\mathbb D$ i.e. ${\mathbb K}/{\mathbb Z}_2$.
In the discretized case for any $N$ one can define a {\it fundamental domain} which is representing the discretized version of ${\rm dS}_2/{\mathbb Z}_2\simeq {\mathbb K}/{\mathbb Z}_2$ properly.
Now unlike ${\rm dS}_2$ the spacetime ${\rm dS}_2/{\mathbb Z}_2$ has one space-like boundary, which clearly corresponds to ${\partial}{\mathbb D}$. Hence the causal diamond structure of ${\rm dS}_2/{\mathbb Z}_2$ can also be regarded as a holographic representation of entanglement patterns of the CFT vacuum state residing on this space-like boundary.
This observation provides a conceptual framework to link our entanglement based considerations to the geometry of $\tilde{\mathcal K}^{N-3}$.
This is the link we are intending to further elaborate on.

\subsection{Factorization and separability}

We have seen that according to (\ref{kezdet}) for each fixed $N$ 
we can gauge away the entanglement entropies $S_{a,a+1}$ at the boundary of the central belt of $\mathbb K$. 
Then we can pose the question where else can we also do that simultaneously? This is precisely the same question posed in \cite{AHST} for planar variables. Let us recapitulate this argument adapted to our entanglement context.

Let us consider a geodesic $N$-gon in $\mathbb D$ and a set $a<b<c<d\in\partial\mathbb D$ of four boundary 
points corresponding to four vertices of this $N$-gon. The point curves of these four vertices give rise to a causal diamond in $\mathbb K/{\mathbb Z}_2$.
Then according to Eq.(\ref{strongsubadditivity})
if we set $S_{a,c}=0$ we cannot set $S_{b,d}$ to zero at the same time without violating strong subadditivity.
Note that the Ryu-Takayanagi geodesics corresponding to such boundary regions  $[a,c]$ and $[b,d]$ are intersecting hence their points in $\mathbb K/{\mathbb Z}_2$ are space-like separated. Then we cannot gauge away the entanglement entropies for any pair of boundary regions represented by space-like separated points in $\mathbb K/{\mathbb Z}_2$. 
In particular, let us fix a boundary region taken together with its representative point in $\mathbb K/{\mathbb Z}_2$.
Then for a collection of boundary regions
represented by points
 lying in the {\it complement of the light cone} of this fixed point in $\mathbb K/{\mathbb Z}_2$ we cannot gauge away the entanglement entropies.
These are just the boundary regions having nontrivial intersection with our fixed boundary region.

\begin{figure}
\centerline{\includegraphics[height=12cm]{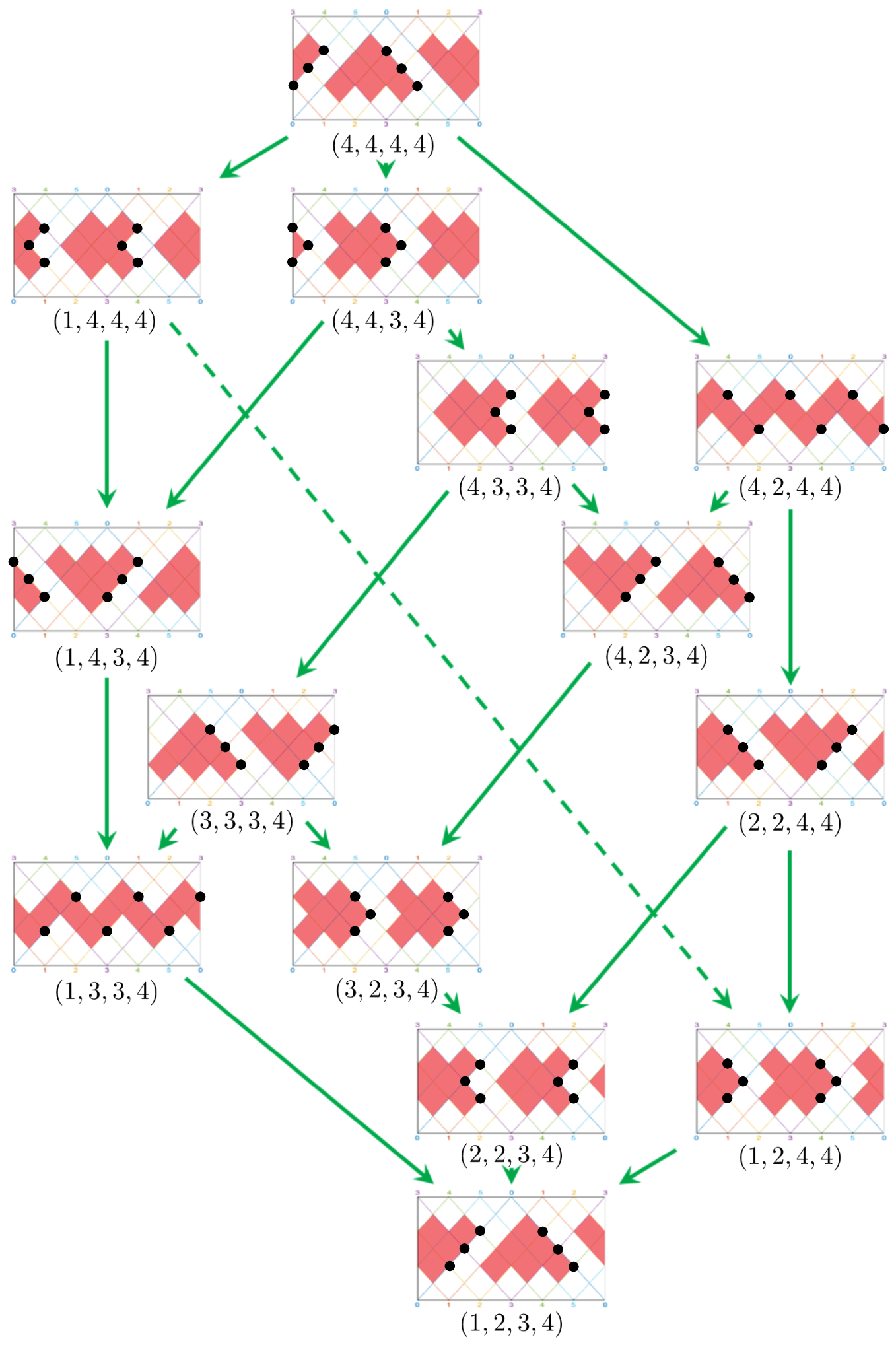}}
\caption{The distribution of triples of points giving rise to the $C_4=14$ causal patterns corresponding to the vertices of the associahedron ${\mathcal K}^3$. Here $C_{N-2}$ is the Catalan number. 
 Such triples of points, indicated by black bullets, are representing the $N=6$ boundary regions where the regularized entanglement entropies can be gauged away simultaneously.
For the six boundary regions the $14$ patterns of entanglement correspond to the $14$ patterns of poles that can appear together in six particle scattering at the tree level. Notice that the duplicated patterns are identified by the M\"obius identification showing up in the definition of elliptic de Sitter space ${\rm dS}_2/{\mathbb Z}_2$.
For the meaning of the labels illustrating the partial Tamari order on this set of patterns see Ref.\cite{LB} and Section IV.C.}
\end{figure}

We have seen that ${\rm dS}_2/{\mathbb Z}_2\simeq \mathbb K/{\mathbb Z}_2$ can be represented by some fundamental domain. In the continuum case an example for a domain of that kind can be choosen as the triangular region inside the past light cone bounded by the point curve of a point. For an illustration of such a domain see the yellow region of Figure 2. of Ref.\cite{Verlinde}. In the discretized case for a fixed $N$ a fundamental domain can be represented by $(N-2)(N-3)/2$ causal diamonds belonging to the central belt. 
In Ref.\cite{LB} for the collection of such diamonds we coined the term {\it causal patterns}. Clearly they can be regarded as $1+1$ dimensional space-times representing ${\rm dS}_2/{\mathbb Z}_2$.
For example for $N=6$ one of the triangular regions of Figure 2. labelled as $(4,4,4,4)$ consisting of six diamonds is a causal pattern of that kind. We emphasize here that for this count we are disregarding $N-2$ extra diamonds that stretch out to the boundary of ${\rm dS}_2/{\mathbb Z}_2$ hence have diverging area.
For fixed $N$ these extra diamonds are needed in order to represent each point in our discretized space-time up to identification of coordinates of Eq.(\ref{uv}) in the form $(u,v)\simeq (v,u)$.
For further clarification on this point and other examples of fundamental domains in the scattering context see Figure 4. of Ref.\cite{AHST}.

Now we can conclude that choosing any pair of non space-like separated points for their representative regions one can always gauge away the regularized entanglement entropies.  
Moreover, one can also do that for space-like separated points that are far enough so that there are no causal diamonds connecting them fitting into the fundamental domain.

In order to illustrate this let us consider the set of causal patterns of the $N=6$ case\cite{LB}.
In Figure 2. one can see the collection of $C_{N-2}=14$ patterns corresponding to the vertices of the associahedron  $\tilde{\mathcal K}^3$. Here $C_{N-2}$ is the Catalan number.
The triples of black bullets are representing those Ryu-Takayanagi geodesics for which the entanglement entropies can be (nontrivially) gauged away. 
As discussed in \cite{LB} the location of these bullets determines the shape of the causal pattern. These patterns are consisting of $(N-2)(N-3)/2$ elementary causal diamonds. Their {\it finite} area labels are fixed by the conditional mutual informations.
We also recall that there is a partial order relation on the space of such patterns called the Tamari order\cite{Tamari}, which can be made explicit by the use of suitable $N-2$-vectors\cite{LB}.

Now from Figure 2. one can see that one can always set the entropies of a pair of non space-like separated points zero, and also space-like separated points that are far enough so that there are no causal diamonds that fit in the fundamental domain representing the space-time. Indeed, the patterns labelled by the vectors $(4,2,4,4)$ and $(1,3,3,4)$ give examples to the latter situation. Although here some of the bullets are space-like separated, however the corresponding causal diamond is too big to fit in the fundamental domain representing our discretized space-time. Hence for these space-like separated points the restrictions provided by strong subadditivity are not effective, hence the entanglement entropies can be made simultaneously zero. 
On the other hand for the pattern labelled by $(4,3,3,4)$ one has pairwise non space-like separated bullets, hence for such points one can have simultaneously vanishing entropies.

Using the elementary causal diamonds inside the pattern labelled by $I_{j,k}\equiv I(b-1b,cc+1\vert bc)$, knowing the boundary values on the right hand side of the domain one can then reconstruct all the $S_{a,b}$ labels for left lying points 
in $\mathbb K/{\mathbb Z}_2$ that we cannot set to zero. Then using the scrunching idea explained in Ref.\cite{AHST} the regions where the entanglement entropies cannot be gauged away, can be scrunched away. What is to be left after this procedure is a spacetime which is factorized. For example for a fundamental domain of the form of a past light cone the factorization yields two past light cones (see Figure 10. of \cite{AHST} for an illustration. Note the flip between their and our notions of space and time, resulting on a $90$ degree rotations of the triangles.) 
This trick can be repeated for any causal pattern fitting into fundamental domains with characteristic curves on the left and right sides identified with the antipodal map.

One can summarize these simple observations 
as follows.
Represent boundary regions, with their reduced density matrices coming from the CFT vacuum, in the asymptotic space-like boundary of ${\rm dS}_2/{\mathbb Z}_2$.
Then the act of gauging away the entanglement entropies,
which is the landmark of separability in entanglement theory, manifests itself in space-time factorization inside a fundamental domain i.e. in ${\rm dS}_2/{\mathbb Z}_2$. 

As stressed in \cite{AHST} for a fixed $N$ a domain to be factorized can be represented by any spacetime domain containing exactly one copy of the $N(N-3)/2$ causal diamonds. 
Hence different spacetime domains can boil down to the same geometric structure of the associahedron $\tilde{\mathcal K}^{N-3}$. Since the number of vertices of $\tilde{\mathcal K}^{N-3}$ is given by the Catalan number $C_{N-2}$
we have the same number of causal patterns in $\mathbb K/{\mathbb Z}_2$ consisting of $(N-2)(N-3)/2$ causal diamonds with their structure determined by the distribution of $N-3$ points. Note however, that the different {\it space-time structures} coming from fundamental domains is much less than $C_{N-2}$. For example by inspection of Figure 2. we see that for $N=6$ we have merely four different characteristic space-time structures. These structures can be captured by the different orientations for the arrows of the $A_{N-3}$ quivers\cite{GenerAssoc}.
In any case the characteristic points of a causal pattern are compatible ones, meaning that the entanglement entropies of their associated boundary regions can be gauged away simultaneously.  

For the $N=6$ case
the distribution of the characteristic points (black bullets of Figure 2.) can also be represented on a particularly choosen pattern. One can choose for example the middle one labelled by $(4,4,3,4)$. Then on this pattern the $14$ different distributions of black bullets coincide with the ones seen on the pictograms of Figure 1. of Ref.\cite{GenerAssoc}. As explained there these pictograms can be obtained from the $A_3$ quiver by an explicit construction. These pictograms can also be used to label the $14$ vertices of the associahedron.

In the scattering context it is well-known that the tree level amplitudes of $N$-particle scattering in the bi-adjoint $\phi^3$ theory have poles when the sum of a subset of external momenta goes on-shell. In this case the residues on these poles factorise into a product of lower level amplitudes. Moreover, these tree-level amplitudes are having $N-3$ poles at a time and the pattern of the set of poles that appear in concert are captured by the structure of our $C_{N-2}$ causal patterns\cite{Nima1,AHST}.
The $N(N-3)/2$ facets of the associahedron ${\mathcal K}^{N-3}$ are labelled by the planar variables $X_{a,b}$ or equivalently the ones of $\tilde{\mathcal K}^{N-3}$ by the nontrivial entanglement entropies $S_{a,b}$. In our  analogy sending these entropies to zero is analogous to having a pole. Moreover, the facets of $\tilde{\mathcal K}^{N-3}$ factorize combinatorially into products of lower dimensional associahedra precisely under simultaneous vanishing of certain subsets of entanglement entropies. 
Hence in our analogy once again we see that the factorization patterns of the scattering amplitudes correspond to the separability properties of the CFT vacuum which in turn correspond to the separable structures of causal patterns in ${\rm dS}_2/{\mathbb Z}_2$.

\subsection{The associahedron as a holographic entanglement polytope}

In the scattering context the ${\mathcal K}^{N-3}$ associahedron structure is determined by the conditions provided by Eqs.(\ref{entropypositive})-(\ref{subadd}) which generalize to the so called $c$-deformed mesh relations well-known from the theory of cluster algebras\cite{GenerAssoc}. These relations combined with the argumentation applied in the subject of positive geometry provide the basis of recent developments connected with the amplituhedron\cite{NimaTrnka}. These mesh relations were applied in a simple manner in \cite{AHST} to show how the causal structures in kinematic space-time yield the structure of the associahedron i.e. the tree-level amplituhedron. Moreover, precisely such relations provide the mathematical basis for generalizations of the ABHY construction for other amplituhedra besides the tree level one\cite{GenerAssoc,AHST}. 

Observing now that the strong subadditivity relation has the appearance of a $c$-deformed mesh relation for the $A_{N-3}$ quiver, and reinterpreting the steps presented in \cite{AHST}, one can immediately conclude that a $\tilde{\mathcal K}^{N-3}$ associahedron structure is also determined by the conditions of Eqs.(\ref{posent}) and (\ref{strongsubadditivity}). In this manner we have yet another physical argument yielding the associahedron obviously and inevitably. However, in arriving at this conclusion this time the physical input provided by the ideas of the subject of holographic entanglement entropy is used.

In order to elaborate on these observations let us consider the $N=6$ case and Figure 2.
In Figure 2. we have different causal patterns consisting of six causal diamonds determined by three characteristic points.
As we change from one full triangulation of $\mathbb D$ by three geodesics to another one by a flip, these characteristic points are changing their positions accordingly.
One can consider the motion of these points within a specially choosen pattern corresponding to a fundamental domain. 

Let us choose this pattern as the triangular one showing up in the middle of the pictogram labelled by $(4,4,4,4)$.
Restrict then first our attention to the three step walk $(4,4,4,4)\to (1,4,4,4)\to (1,2,4,4)\to (1,2,3,4)$ in this triangular region. We observe that the resulting walk is featuring only points located either on the right or on the left hand side of the triangular region shown in Figure 3. Of course generally under a walk the red points are moving from right to left via also featuring intermediate points like the ones in Figure 4.\footnote{Under a clockwise rotation of the triangular pattern by $90$ degrees this walk corresponds to the upward directed motion i.e. the "time evolution" of Ref.\cite{AHST}.} 

Recall also that at each point of the triangular pattern  we have an entanglement entropy variable $S_{ab}\geq 0$ taken from the {\it physical branch}. Specially, the black bullets of Figure 3. correspond to the gauge fixed entropies $S_{12}=S_{23}=S_{34}=S_{05}=0$. 
Notice also that each causal diamond also carries a pre assigned label of conditional mutual information $I_{ab}\geq 0$.
For example by virtue of Eqs.(\ref{areas}) and (\ref{zsolt}) for a regular hexagon one can show that
\begin{equation*}
I_{04}=I_{02}=I_{13}=I_{24}=\frac{\mathfrak{c}}{3}\log\frac{3}{2},\quad I_{03}=I_{14}=\frac{\mathfrak{c}}{3}\log\frac{4}{3}.
\end{equation*}
It is important to realize at this stage that the constraints $S_{ab}\geq 0$ and $I_{ab}\geq 0$ in the entanglement context are dictated by {\it physical principles}, namely the nonnegativity of the entropy and strong subadditivity.
Mathematically these constraints are analogous to the ones of $X_{ab}\geq 0$ and $c_{ab}\geq 0$ needed for invoking the magic of {\it positive geometry} as emphasized in Ref.\cite{AHST} in the scattering context.
However, unlike in the kinematic space-time considerations of \cite{AHST} where these constraints were injected by hand, here they are following from the physics of entanglement.
In our $N=6$ example apart from these constraints dictated by positive geometry, we have freedom in choosing the remaining $9$ entropy labels.

\begin{figure}
\centerline{\includegraphics[height=2cm]{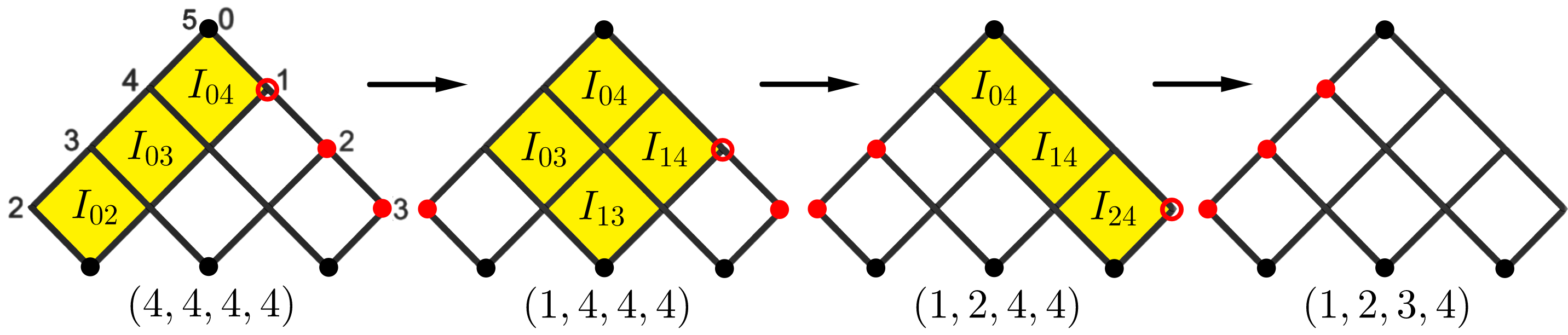}}
\caption{The three step walk from the configuration $(4,4,4,4)$ to the one $(1,2,3,4)$ giving rise to a cuboid ${\mathcal C}^3$ in the three dimensional orthant spanned by the entropies $S_{15}, S_{25},S_{35}$. }
\end{figure}

\begin{figure}
\centerline{\includegraphics[height=2cm]{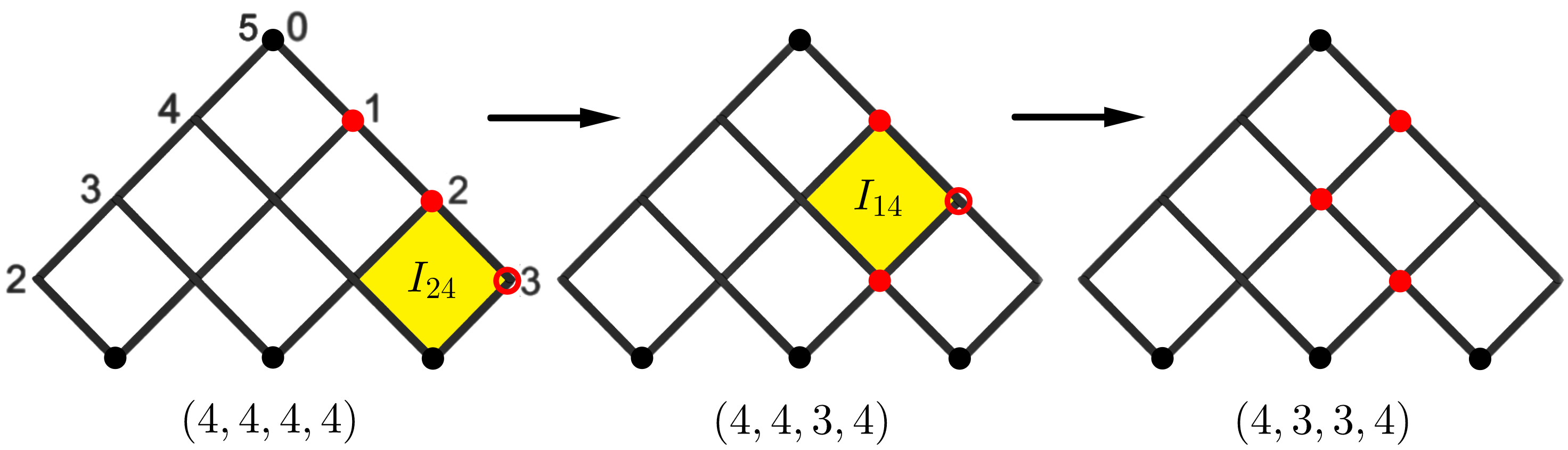}}
\caption{A two step walk giving rise to the extra constraints of Eqs.(\ref{extra}) needed for carving out the associahedron
$\tilde{\mathcal K}^{3}$
from the cuboid ${\mathcal C}^{3}$ created by the walk of Figure 3.}
\end{figure}

Now our special walk starts with choosing a point, in our case this is the first one located on the grid point $15$ denoted by a red circle.
Then the first step of the walk is effected by moving this point $15$ to its new position $02$ determined by completing the causal diamond located to the {\it left} (see the yellow region) of the red circle.
Clearly this step is associated with the strong subadditivity constraint
\begin{equation}
I_{02}+I_{03}+I_{04}=S_{02}+S_{15}-S_{12}-S_{05}=S_{02}+S_{15}\geq 0
\end{equation}
where in the left hand side we have used the additivity of conditional mutual information.
Repeating the same steps with the second and third points, i.e. the ones located at the grid points $25$ and $35$ one obtains the inequalities 
\begin{equation*}
S_{02}=I_{02}+I_{03}+I_{04}-S_{15}\geq 0
\end{equation*}
\begin{equation*}
S_{03}=I_{03}+I_{04}+I_{13}+I_{14}-S_{25}\geq 0
\end{equation*}
\begin{equation*}
S_{04}=I_{04}+I_{14}+I_{24}-S_{35}\geq 0.
\end{equation*}
These inequalities are of the form $C_k\geq S_{k5}$ with $k=1,2,3$ and $C_k\geq 0$, and carve out a three dimensional {\it cuboid} ${\mathcal C}^3$ in the orthant defined by $S_{k5}\geq 0$.
For example, for the regular hexagon case (six equal boundary regions) we have $C_1=C_3=\frac{\mathfrak{c}}{3}\log3$ and
$C_2=\frac{\mathfrak{c}}{3}\log4$.

The associahedron ${\mathcal K}^3$ is carved out from ${\mathcal C}^3$ by employing extra constraints coming from steps also featuring intermediate points of our triangular causal pattern.
For example by inspection of Figure 4. one can check that from this two step process one obtains
\begin{equation*}
I_{24}=S_{24}+S_{35}-S_{25},\quad
I_{14}=S_{14}+S_{25}-S_{24}-S_{15}.
\end{equation*}
The first of these equations, and the sum of the two gives the constraints
\begin{equation}
S_{35}\leq I_{24}+S_{25},\quad S_{35}\leq I_{14}+I_{24}+S_{15}
\label{extra}
\end{equation}
giving rise to two planes cutting ${\mathcal C}^3$ further down. Continuing the example of Figure 4. it is easy to obtain the remaining constraint to carve out ${\mathcal K}^3$
which is
\begin{equation*}
S_{25}\leq I_{13}+ I_{14}+S_{15}.
\end{equation*}
The result can be seen in Figure 5.

\begin{figure}
\centerline{\includegraphics[height=8cm]{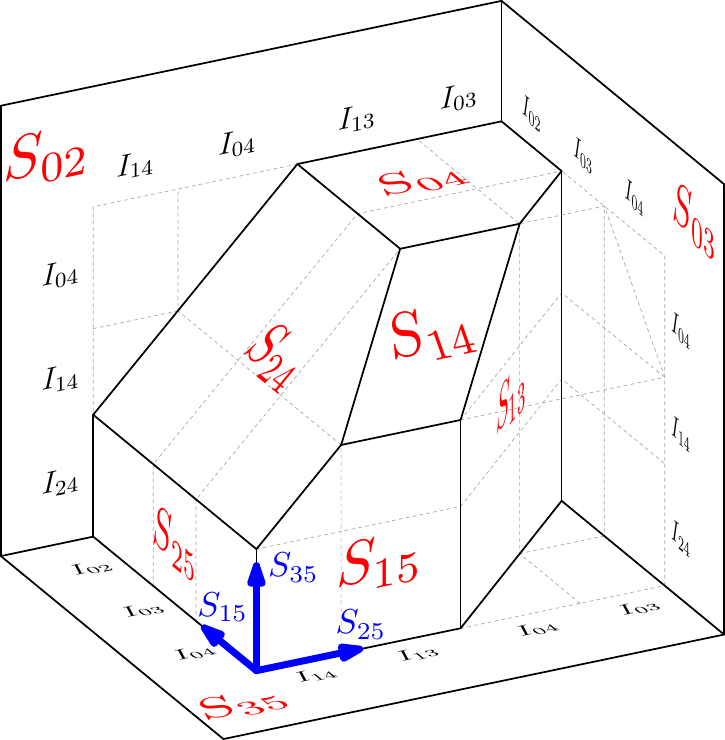}}
\caption{The $\tilde{\mathcal K}^3$ associahedron.}
\end{figure}

As an explicit example one can consider the regular hexagon case with
$A=\frac{\mathfrak{c}}{3}\log2$, $B=\frac{\mathfrak{c}}{3}\log3$ and the associahedron is defined by the inequalities
\begin{equation*}
x\leq B,\quad y\leq 2A,\quad z\leq B
\end{equation*}
\begin{equation*}
z-y\leq B-A,\quad z-x\leq A,\quad y-x\leq A
\end{equation*}
where $(x,y,z)=(S_{15},S_{25},S_{35})$.

In the picture based on a walk in a causal pattern, it is easy to understand the meaning of the coordinates of the Tamari lattice\cite{LB} labelling the $C_{N-2}$ causal patterns, rendering the partial order structure explicit. We illustrate this for the $N=6$ case.
First of all the fourth coordinate is always $4$ hence we merely have to account for the numbers occurring in the first three slots. The {\it label of the slot} corresponds to the label of the red circle which is to be moved. This slot label also equals the label of the right moving light cone coordinate line, the line which the circle occupies before it moves. 
The {\it number in the slot} is the number of walk steps which is to be taken by the circle down on that light cone coordinate line when moving to the final position.
Under the walk of Figure 3. during the first move the corresponding circle took three, in the second two, and in the third just one step down the corresponding lines, before switching to the left moving light cone line orienting the circle to its final position.
Under the walk of Figure 4. we have taken just {\it single step} moves.
One can choose a walk starting from $(4,4,4,4)$ and arriving to $(1,2,3,4)$ consisting of just single step walks.
Indeed, continuing the walk of Figure 4. via taking single steps we obtain an example of this walk (see also Figure 13. of Ref.\cite{AHST}). The advantage of such a walk is that in this case the unique walk 
\begin{equation*}
(4,4,4,4)\to (4,4,3,4)\to(4,3,3,4)\to(3,3,3,4)\to
\end{equation*}
\begin{equation*}
(3,2,3,4)\to(2,2,3,4)\to(1,2,3,4).
\end{equation*}
generates all the relevant inequalities needed for carving out ${\mathcal K}^3$. 

Our considerations are easy to generalize. We start with a causal pattern of triangular shape consisting of $(N-2)(N-3)/2$ causal diamonds, with its upmost grid point $(N-1,0)$. We start with $N-3$ points located on the right hand side of the triangular region labelled by the $N-2$ vector $(N-2,N-2,\dots,N-2)$. The walk terminates when the $N$ points are reaching the left hand side of the triangular region, a configuration labelled by $(1,2,3,\dots,N-2)$. As the walk enfolds the first $N-3$ coordinates of our $N-2$ vector can then be generated recursively by an analogue of the process described for $N=6$.
In particular an $N-3$ step walk analogous to the one of Figure 3. carves out a cuboid ${\mathcal C}^{N-3}$ and the relations connected to intermediate points cut it further down to the associahedron ${\mathcal K}^{N-3}$.
There is again a unique single step walking process generating all the inequalities neeeded for the cutting down process creating ${\mathcal K}^{N-3}$.
Notice that this unique walk can be nicely encapsulated in an underlying  walk described in \cite{AHST,GenerAssoc} on the quiver associated with the $A_{N-3}$ Dynkin diagram.

The upshot of these considerations is that to the ${\rm CFT}_2$ vacuum state living on the boundary of the static slice $\mathbb D$ of ${\rm AdS}_3$ one can associate the polytope $\tilde{\mathcal K}^{N-3}$ living in the $N-3$ dimensional orthant spanned by the entanglement entropies $S_{k,N-1}$ with $k=1,2,\dots N-3$.
The parameters of this associahedron are determinded by certain sums of the $I_{ab}$ gauge invariant measures of entanglement.
The codimension $d$ boundaries of $\tilde{\mathcal K}^{N-3}$ correspond to the combinatorial possibilities of $d$ entanglement entropies from the set $S_{ab}$ ($0\leq a<b\leq N-1$ with $a,b$ non adjacent) simultaneously conspiring to vanish under the guidance of a gauge degree of freedom.  
Since 
$\tilde{\mathcal K}^{N-3}$ is encoding data on holographic entanglement patterns of the CFT vacuum state it can be regarded as a holographic entanglement polytope.

\subsection{Worldsheet associahedron}

Now we would like to elaborate on the connection between the associahedra $\tilde{\mathcal K}^{N-3}$ and ${\mathcal K}^{N-3}$ of Section III.
Notice that in the literature the associahedron has already appeared in two different contexts.
The first one is the kinematic
associahedron which first appeared within the context of bi-adjoint scalar amlitudes as a twisted cycle in Ref.\cite{Mizera}.
This object is also known as
the ABHY associahedron proposed in \cite{Nima1} which is ${\mathcal K}^{N-3}$.
However, in the very same paper it has also been clarified (see references in that paper) that the associahedron is at the same time showing up in connection with the moduli space $\mathcal{M}_{0,N}$ of $N$ punctures on the Riemann sphere, a result well-known to string theorists\cite{Devadoss,Hanson,Cruz}. 
In the literature this associahedron is called the "world-sheet associahedron".
The reader might have already noticed that it is this latter worldsheet associahedron which is to be related to our $\tilde{\mathcal K}^{N-3}$.

Let us then finally examine this obvious guess.
When considering the worldsheet associahedron one should consider
the positive part of the real part of $\mathcal{M}_{0,N}$, which is the positive open string moduli space $\mathcal{M}^+_{0,N}(\mathbb R)$.
Explicitely we have
\begin{equation}
\mathcal{M}^+_{0,N}(\mathbb R)=\{\xi_0<\xi_1<\cdots\xi_{N-1}\}/SL(2,{\mathbb R})
\end{equation}
which is our setup of $N$ ordered points on $\partial{\mathbb U}$.
Then the world-sheet associahedron\cite{Nima1} is the Deligne-Mumford-Knudsen compactified (see Ref.\cite{Cruz} and references therein)  positive moduli space $\overline{\mathcal{M}}^+_{0,N}$.

On the other hand our construction of 
$\tilde{\mathcal K}^{N-3}$
 was based on the analogy of Eq.(\ref{alapmegfeleles}) and the use of the gauge degree of freedom which enabled us to gauge away the entanglement entropies for the sides of a geodesic $N$-gon.
This gauge degree of freedom was based on the freedom of choosing the pair of horocycles at the corresponding boundary points. Now horocycles in the static setting are mathematically described by future directed light-like vectors\cite{Penner,Pennerbook,Levay} i.e. elements of ${\mathcal L}^+$ in ${\mathbb R}^{2,1}$. For the upper sheet of the double sheeted hyperboloid $\mathbb H$, providing our model for the bulk $\mathbb D$, horocycles $h_a$ and $h_b$ are special curves consisting of points  $y\in \mathbb H$ defined by a pair of future directed light-like vectors $n_a,n_b\in{\mathcal L}^+$ such that\footnote{
Here by an abuse of notation we have used $\cdot$ for the inner product in ${\mathbb R}^{2,1}$ not to be confused with the same symbol showing up in Eq.(\ref{Mandelstam}). For the occurrence of the factor $1/\sqrt{2}$ see the book of Penner\cite{Pennerbook}.}
\begin{equation}
y\cdot n=-1/\sqrt{2}, \quad y\cdot y=-1, \quad  n\cdot n=0.
\label{horo}
\end{equation}

The correspondence between the set of horocycles and the set of points in ${\mathcal L}^+$ is one to one.

Now for the lambda length we have the alternative formula\cite{Penner,Pennerbook}
\begin{equation}
\lambda_{a,b}\equiv\lambda(h_a,h_b)= \sqrt{-n_a\cdot n_b}.
\label{lambdapenner}
\end{equation}
Then for the regularized entropies we have
\begin{equation}
S_{a,b}=\frac{\mathfrak{c}}{3}\log\lambda_{a,b}.
\label{horoentro}
\end{equation}

Represent now these light-like vectors 
$n_a,n_b$ associated with the boundary points $a$ and $b$
as real $2\times 2$ symmetric matrices of the form $\hat{n}_a=\sqrt{2}{\bf a}{\bf a}^T$ and $\hat{n}_b=\sqrt{2}{\bf b}{\bf b}^T$.  Hence we can use instead of them the spinors ${\bf a},{\bf b}\in {\mathbb R}^2$. Then in this picture the Minkowsi inner product is represented as $-n_a\cdot n_b\equiv\langle{\bf a}{\bf b}\rangle^2\equiv (a_1b_2-a_2b_1)^2$ with $\langle{\bf a}{\bf b}\rangle>0$.
Hence the regularized entanglement entropy can also be expressed in terms of the spinors $\bf a$ and $\bf b$
as
\begin{equation}
S_{a,b}=\frac{\mathfrak{c}}{3}\log\langle{\bf a}{\bf b}\rangle.
\end{equation}

For $N$ boundary points we have $N$ such spinors. This set of $N$ two component vectors can be arranged in a $2\times N$ matrix representing an element of the positive Grassmannian of two-planes in ${\mathbb R}^N$ i.e. $Gr^+(2,N)$. Then the quantities $\langle{\bf a}{\bf b}\rangle>0$ showing up in the lambda lengths, related to the $S_{ab}$ regularized entropies, are positive minors corresponding to the usual Pl\"ucker coordinates of the positive Grassmannian. 
In this picture the gauge degree of freedom of choosing the $N$ diameters $\Delta_a,\Delta_b,\dots$ of the $N$ horocycles for the $N$ boundary points (see Figure 1.) corresponds to the torus action ${\mathbb R}^N_{>0}$ on $Gr^+(2,N)$. This group is acting on each of the columns of the $2\times N$ matrix. Then ${\mathcal{M}}^+_{0,N}$ can also be regarded as  $Gr^+(2,N)$ modded out by this action.

Choosing now a special partition of the boundary into $N$ regions via fixing an ordered set of $N$ points taken together with a special set of $N$ diameters for the horocycles, amounts to a gauge fixing procedure for the group $SL(2,{\mathbb R})\times GL^+(1,\mathbb R)^N$.
As a result of these considerations we see that our gauging away of the entanglement entropies for the edges of the geodesic $N$-gon corresponds to a particular choice of gauge for the torus action, which is rendering the lambda lengths $\lambda_{a,a+1}=1$.

Recall now the well-known fact that the open superstring amplitudes are defined as integrals over ${\mathcal{M}}^+_{0,N}$ in the form\cite{Nima3}
\begin{equation}
{\bf I}_N(\{s\})=(\alpha^{\prime})^{N-3}\int_{{\mathcal{M}}^+_{0,N}}\frac{d^{N-3}}{\xi_{0,1}\cdots\xi_{N-1,0}}\prod_{a<b}
(\xi_{a,b})^{\alpha^{\prime}s_{a,b}}
\label{opensuper}
\end{equation}
where $\xi_{a,b}=\xi_b-\xi_a$.
After recalling Eq.(\ref{subadd}) an alternative form for this integral can be given by rewriting the Koba-Nielsen factor\cite{KN}
in the form\cite{Nima2}
\begin{equation}
\prod_{a,b}{\langle{\bf a}{\bf b}\rangle}^{-\alpha^{\prime}c_{a,b}}=\prod_{i,j}u_{i,j}^{\alpha^{\prime}X_{i,j}}
\label{kobak}
\end{equation}
where
\begin{equation}
u_{i,j}=U_{i-1,j-1}
\label{jaj}
\end{equation}
of Eq.(\ref{uvar}). Moreover, since on the 
right hand side the torus action has been modded out there are only $N(N-3)/2$ independent $s_{ab}$ converted to planar variables $X_{i,j}$.
In this new description\cite{Nima3} the measure of integration is given by
\begin{equation}
\Omega({\mathcal U}^+_N)=\prod_{k=1}^{N-3}d\log\frac{U_k}{1-U_k}
\label{parketaylor}
\end{equation}
where $U_k=U_{N-1,k}$ correspond to the seed variables of Eq.(\ref{seedset}) that by virtue of Eqs.(\ref{fs1})-(\ref{fs2}) determining all of the $u_{i,j}$ showing up in the right and side of (\ref{kobak}).

Now we see that formally the world sheet associahedron i.e. the moduli space 
$\overline{\mathcal{M}}^+_{0,N}$ can also be regarded as the entanglement polytope of the CFT vacuum in the static setting of ${\rm AdS}_3/{\rm CFT}_2$. Then by virtue of Eq.(\ref{osszefuggesek}) and (\ref{jaj}) integration on the moduli space with respect to the (\ref{parketaylor}) Parke-Taylor form corresponds to integration with respect to  different $U_k$ seed patterns of entanglement. 
The advantage of this $u_{ab}\sim -\log I_{ab}$ presentation is that the open string integral obtained in this way is completely gauge invariant (the conditional mutual informations are not depending on the horocycles).
Moreover, our analogy combined with Eq.(\ref{kobak}) also suggests a formula
\begin{equation}
e^{-\frac{3\alpha^{\prime}}{\mathfrak{c}}\sum_{a,b}c_{a,b}S_{a,b}}
=e^{-\frac{3\alpha^{\prime}}{\mathfrak{c}}\sum_{i,j}I_{i-1,j-1}X_{i,j}}.
\label{entkobak}
\end{equation}
showing that a Koba-Nielsen-like factor naturally relates the quantities $(S_{a,b},I_{a,b})$ and $(X_{a,b},c_{a,b})$ hence the associahedra ${\mathcal K}^{N-3}$ and $\tilde{\mathcal K}^{N-3}$.

Interestingly the analogous relationship in the $\overline{\mathcal{M}}^+_{N,0}$-${\mathcal K}^{N-3}$ context  in the interior of the polytope leads us to the diffeomorphism discussed in Ref.\cite{Nima1} relating the worlds-sheet associahedron and the kinematic one via the scattering equations\cite{scatt}. As is known the scattering equations are saddle point equations (the vanishing of $d\log$ of the Koba-Nielsen factor) of the (\ref{opensuper}) integral in the limit $\alpha^{\prime}\to \infty$. 
These equations have the implicit form $\sum_{a<b}c_{a,b}dS_{a,b}=0$.
As far as the boundaries are concerned it is also known that every boundary of
$\overline{\mathcal M}_{0,N}$ of any codimension is mapped to the corresponding one of ${\mathcal K}^{N-3}$.
Technically this means that an $u_{a,b}=0$  boundary is mapped into an $X_{a,b}=0$ one\cite{Nima1,Nima3}.

\section{Conclusions}

According to the ideas of Ref.\cite{AHST} structures of scattering amplitudes that have been hidden in the usual approach of Feynman diagram treatment building on the principles of unitarity and locality should be made exlicit.
In particular the patterns of compatible poles of tree level amlitudes appearing together are structures of that kind.
It has turned out that for $N$-paricle scattering in bi-adjoint $\phi^3$ theory such compatible poles correspond to the vertices of the associahedron ${\mathcal K}^{N-3}$. Moreover, the factorization properties of the amplitudes are geometrized by the factorization properties of the associahedron into lower dimensional associahedra.

In this paper we have developed an analogy between such patterns of poles, and the patterns of entanglement for $N$ subsystems of the CFT vacuum in the static ${\rm AdS}_3/{\rm CFT}_2$ setup 
studied in Ref.\cite{LB}.
Both of these patterns are automatically taken care by the geometric properties of the associahedron. We argued that for patterns of entanglement the latter associahedron can be regarded as a holographic entanglement polytope.
In arriving at this picture we emphasized the importance of the physical principles of the positivity of the physical branch of the entropy $S_{ab}\geq 0$, and strong subadditivity $I_{ab}\geq 0$.
Indeed, these principles automatically ensure the constraints needed for the usual magic of positive geometry, familiar from the scattering context, to work.

In our treatment it turned out that the factorization properties of amplitudes in the scattering picture are analogous to the separability properties of the CFT vacuum.
Moreover, the separability properties of the CFT vacuum manifest themselves in separability properties of the kinematic space ${\mathbb K}/{\mathbb Z}_2$. This space can be regarded as a space-time in it own right : elliptic de Sitter space. 
Hence in this picture different patterns of entanglement of the vacuum are geometrized to different causal patterns serving as representatives for a space-time: ${\mathbb K}/{\mathbb Z}_2$.
As we change $N$ (the number of subsystems the boundary is partitioned into) the geometric structure of the associahedron, as a one organized into lower dimensional associahedra, encodes how this space-time is glued together from patterns of entanglement in a holographic manner.

We also elaborated on the interesting dynamics on the space of causal patterns effected by certain "flip" transitions, having their roots in quiver mutations of an underlying $A_{N-3}$ cluster algebra\cite{Levay,LB}.
We have shown how this dynamics is enfolding as a walk of $N-3$ particles on a causal pattern. We pointed out that this walk is directly related to the well-known walk discussed in the scattering context\cite{AHST}.

Finally we observed that the associahedron as a holographic entanglement polytope for the CFT vacuum can be formally identified with the world sheet associahedron of the moduli space $\overline{\mathcal{M}}^+_{0,N}$. Moreover, certain gauge degrees of freedom used under this formal identification acquire a new meaning: the gauge choice for the torus action rendering the lambda lengths to unity, correspond to the gauging away of the entanglement entropies for the edges of the bulk geodesic N-gon.

Are there any other holographic entanglement polytopes related to cluster algebras?
Are there further deeper connections between entanglement polytopes, moduli spaces, and scattering theory?
Clearly the ABHY construction can be generalized which results in generalized associahedra.
Moreover, these objects are related to a generalization of the tree level situation to one-loop tadpole diagrams, and one loop amplitudes.
Our speculative correspondence then suggests that an obvious guess in the holographic context is the one
that
generalized associahedra should somehow correspond to special  multiboundary wormhole\cite{Levay,LB} configurations. It is known that these objects can alternatively be described as special fundamental domains in kinematic space\cite{Zhang,Zukowski}.
In particular in Figure 16. of\cite{AHST} a kinematic spacetime region for the $D_4$ quiver appeared.
In this situation the authors have a situation where $X_{a,b}\neq X_{b,a}$. Naively one would think that this is the analogue of $S_{a,b}\neq S_{b,a}$ which suggests a holographic description in terms of a {\it mixed state}.
In order to move from the level of speculations further mathematical elaboration is needed. We are intending to explore such ideas in future work.

\section{Acknowledgement}

This work was supported by the National Research Development and Innovation Office of Hungary within the Quantum Technology National Excellence Program (Project No. 2017-1.2.1-NKP-2017-0001). We would like to thank the help of Bercel Boldis and Zsolt Szab\'o with our Figures.

\end{document}